\documentclass[12pt]{article}

\usepackage{epsfig,amssymb,euscript}
\usepackage{amsmath}

\numberwithin{equation}{section}

\newcommand{\dif}{\textrm{d}}
\newcommand{\me}{\textrm{e}}

\newcommand{\Vol}{{\rm Vol}\,}

\newcommand{\bbR}{\mathbb{R}}
\newcommand{\bbC}{\mathbb{C}}

\newcommand{\ia}{\alpha}          % (alpha) index
\newcommand{\ib}{\beta}           % (beta) index
\newcommand{\ic}{\gamma}          % (gamma) index
\newcommand{\id}{\delta}          % (delta) index

\newcommand{\lk}{l}               % curvature of sigma (+1,0,-1)

\begin{document}

%%%%%%%%%%%%%%%%%%%%%%%%%%%%%%%%%%%%%%%%%%%%%%%%%%%%%%%%%%%%%%%%%%

\baselineskip 18pt

%%%%%%%%%%%%%%%% title page %%%%%%%%%%%%%%%%%%%%%%%%%%%%%%%%%%%%

\begin{titlepage}
\vfill
\begin{flushright}
QMUL-PH-01-04\\
hep-th/0105250\\
\end{flushright}

\vfill

\begin{center}
\baselineskip=16pt
{\Large\bf Membranes Wrapped on}
\\
{\Large\bf Holomorphic Curves}
\vskip 10.mm
{Jerome P. Gauntlett$^{1}$, ~Nakwoo Kim$^{2}$, Stathis Pakis$^3$
and Daniel Waldram$^{4}$}\\
\vskip 1cm
%\vfill
{\small\it
Department of Physics\\
Queen Mary, University of London\\
Mile End Rd, London E1 4NS, UK}\\
\vspace{6pt}
\end{center}
\vfill
\par
\begin{center}
{\bf ABSTRACT}
\end{center}
\begin{quote}
We construct supergravity solutions dual to the twisted field 
theories arising when M-theory membranes wrap holomorphic curves 
in Calabi--Yau $n$-folds. The solutions are constructed in an Abelian
truncation of maximal $D=4$ gauged supergravity and then uplifted to $D=11$.
For four-folds and five-folds we find new smooth AdS/CFT examples 
and for all cases we analyse the nature of the singularities that arise.
Our results provide an interpretation of certain charged
topological AdS black holes. We also present the generalised calibration 
two-forms for the solutions.
\vfill
\vskip 5mm
\hrule width 5.cm
\vskip 5mm
{\small
\noindent $^1$ E-mail: j.p.gauntlett@qmw.ac.uk \\
\noindent $^2$ E-mail: n.kim@qmw.ac.uk \\
\noindent $^3$ E-mail: s.pakis@qmw.ac.uk \\
\noindent $^4$ E-mail: d.j.waldram@qmw.ac.uk \\
}
\end{quote}
\end{titlepage}

%%%%%%%%%%%%%%%%%%%%%%%%%%%%%%%%%%%%%%%%%%%%%%%%%%%%%%%%%%%%%%%%%%

\section{Introduction}

An interesting generalisation of the AdS/CFT correspondence \cite{mald}
is the construction of supergravity solutions dual to the field
theories that arise on branes wrapping supersymmetric cycles.
To preserve supersymmetry it is necessary for the field theory
to be twisted in the sense that there is an identification of the
spin connection on the cycle with certain external R-symmetry gauge
fields \cite{bvs}. It was argued in \cite{malnun} that this implies
that dual supergravity solutions can be found in the appropriate
gauged supergravity and then, ideally, uplifted to ten or eleven
dimensions. This two-step approach enables one to find highly
non-trivial supergravity solutions and has been further developed in a
number of papers~\cite{malnuntwo,agk,no,gkw,nunpst,ednun,st,mn}. Here
we will extend these investigations by considering the theories which
arise when M-theory membranes wrap two-cycles in Calabi--Yau two-,
three-, four- and five--folds. 

For unwrapped membranes recall that there is a decoupling limit
in which one obtains a $D=3$ $N=8$ SCFT that is dual to
M-theory on $AdS_4\times S^7$, with the $SO(8)$ isometries of the
seven-sphere corresponding to the $SO(8)_R$ R-symmetries of the
SCFT~\cite{mald}. When a membrane wraps a two-cycle $\Sigma$ in a
Calabi--Yau $n$-fold it will preserve some supersymmetry if the two-cycle
is holomorphic. The relevant twisting mentioned above then depends on
the structure of the normal bundle of the cycle in the $n$-fold,
as we shall discuss. The twistings we shall consider only involve
the gauge fields of the maximal Abelian subgroup $U(1)^4$ of $SO(8)_R$.
As a consequence we are able to construct the solutions in
a $U(1)^4$ truncated version of $D=4$ $N=8$ gauged supergravity
\cite{duffliu} and then use the results of \cite{duff10} to uplift to
find $D=11$ solutions. 

The gauged supergravity solutions have an $AdS_4$-type region,
specified more precisely below, that describes the UV physics
of the decoupled $D=3$ twisted field theory arising on the wrapped membranes.
The appropriate decoupling limit involves letting $l_{pl}\to 0$ while
keeping the volume of the cycle fixed and implies that
only the local geometry of the holomorphic curve in the
Calabi--Yau manifold is relevant \cite{malnun}. In the IR, at energies
small compared to the energy scale set by the inverse size of the
two-cycle, the theory reduces to a one-dimensional supersymmetric
quantum mechanics. The solutions we obtain describe the flow from the
dual $AdS_4$-type UV region to the gravity dual descriptions of the IR
physics. For twistings corresponding to holomorphic curves in
Calabi--Yau four- and five-folds, we find an IR fixed point with
geometry $AdS_2\times\Sigma$ when the curvature of the two-cycle 
$\Sigma$ is negative. Lifted to $D=11$, these solutions are dual to a
superconformal quantum mechanics and thus provide new AdS/CFT
examples. Interestingly, for the five-fold case, the full lifted
solution describing the flow from the UV, is an embedding into eleven
dimensions of the supersymmetric ``topological'' charged $D=4$ $AdS$-black
hole of~\cite{caldklemm}. Our analysis thus indicates the proper
interpretation of this solution. In addition we are able to lift
the rotating generalisation of the $D=4$ black hole discussed 
in~\cite{caldklemm} to obtain a new supersymmetric $D=11$ solution
that describes supersymmetric waves on the wrapped membranes.

Since the $D=11$ solutions describe the geometry arising when membranes 
wrap the two-cycle $\Sigma$ and also preserve supersymmetry, on
general grounds we would expect that a probe membrane wrapping
the same cycle $\Sigma$ will also be supersymmetric. Given that our
backgrounds are static and  have non-vanishing four-form, this means
that our solutions should admit generalised K\"{a}hler calibration
two-forms \cite{gpt}, which we shall explicitly present. 

The solutions presented here for wrapped membranes and in related work
for other wrapped branes go well beyond the intersecting brane
solutions found using the ``harmonic function rule''
\cite{paptown,tseyt,gkt} or the ``generalised harmonic function rule''
\cite{tseyttwo,ggpt}. In \cite{FS1,FS4} a set of more general BPS
equations were derived for certain smoothed intersections of fivebranes
corresponding to fivebranes wrapped on Riemann surfaces and some
solutions were found. Further solutions were presented in \cite{FS2,FS3}. 
This approach was
generalised to obtain BPS equations for fivebranes wrapped on
K\"{a}hler four-cycles in six dimensions and also to membranes wrapped
on Riemann surfaces in \cite{cekt} where the connection with
generalised calibrations was exploited. In order to clarify the
connection between this alternative approach with ours we will present
a change of coordinates that recasts our solutions into the form
considered in \cite{cekt}. Similar coordinate transformations also
exist for the wrapped fivebrane solutions constructed in \cite{gkw}. 

The plan of the rest of the paper is as follows. In the next section
we recall the Abelian truncation of $N=8$ gauged supergravity and
how it relates to $D=11$ supergravity. We then present and analyse the
BPS equations for membranes wrapping two-cycles in Calabi--Yau two-,
three-, four- and five-folds. Section 4 presents the generalised
K\"{a}hler calibration and the coordinate transformations  mentioned in
the last paragraph. Section 5 concludes.

%%%%%%%%%%%%%%%%%%%%%%%%%%%%%%%%%%%%%%%%

\section{$S^7$ reduction and $U(1)^4$ gauged supergravity}
\label{sec:sugra}

The $S^7$ reduction of eleven-dimensional
supergravity gives rise to $d=4$, $N=8$ gauged $SO(8)$
supergravity. We will use the reduction ansatz presented
in~\cite{duff10} which retains only $U(1)^4$ gauge fields.
The eleven-dimensional metric is given by
\begin{equation}
   \dif s^2 = {\Delta}^{2/3} \dif s_4^2
      + {2\over e^2}{\Delta}^{-1/3}\sum_\ia X_\ia^{-1} \left(
           \dif\mu_\ia^2 + \mu_\ia^2\left(\dif\phi_\ia+2eA_\ia\right)^2
           \right) ,
\label{eq:g11}
\end{equation}
where $\dif s_4^2$ is the four-dimensional metric, $X_\ia$ and $A_\ia$, with
$\ia=1,\dots,4$, are scalars and one-forms, respectively, on the
four-dimensional space, and ${\Delta}=\sum_\ia
X_\ia\mu_\ia^2$. The coordinates $\mu_\ia$ and $\phi_\ia$ parametrise
a seven-sphere: the $\mu_\ia$ are
constrained to satisfy $\sum_\ia\mu_\ia^2=1$ and $0\leq\phi_\ia<2\pi$
are angles. The four scalar fields $X_\ia$, satisfying $X_1 X_2 X_3 X_4=1$,
deform the round sphere metric generically
breaking the $SO(8)$ symmetry to $U(1)^4$ parametrised
by rotations in the four angles $\phi_\ia$. In addition
these directions are twisted by the four $U(1)$ gauge fields $A_\ia$.

The four-form field strength is given in terms of the same fields via
\begin{multline}
   G = \sqrt{2}e
          \sum_\ia\left(X_\ia^2\mu_\ia^2-{\Delta}X_\ia\right)\epsilon_4
       - \frac{1}{\sqrt{2}e}\sum_\ia X_\ia^{-1} * \dif X_\ia \wedge
          \dif\mu_\ia^2 \\ 
       - \frac{2\sqrt{2}}{e^2}\sum_\ia X_\ia^{-2} \dif\mu_\ia^2 \wedge
          \left(\dif\phi_\ia+2eA_\ia\right) \wedge * F_\ia ,
\label{eq:G4}
\end{multline}
where $*$ is the Hodge dual operator on
the four-dimensional space and $\epsilon_4$ is the corresponding
volume form.

Reducing with this ansatz leads to a four-dimensional theory with
bosonic action
\begin{equation}
   \mathcal{L} = \frac{1}{2\kappa^2} \sqrt{-g} \left[
       R - \frac{1}{2}(\partial\vec{\phi})^2
       - 2 \sum_\ia \me^{\vec{a}_\ia\cdot\vec{\phi}} F_\ia^2
       - V \right] ,
\label{eq:S}
\end{equation}
where
\begin{equation}
   V = -4e^2 \left( \cosh\phi_{12} + \cosh\phi_{13}
      + \cosh\phi_{14} \right) .
\label{eq:V}
\end{equation}
Here we have introduced a new basis for the constrained scalar fields
$X_\ia$ in terms of a vector of scalar fields $\phi_{\ia\ib}$
\begin{equation}
   \vec{\phi} = (\phi_{12},\phi_{13},\phi_{14}) ,
\label{eq:phi}
\end{equation}
where under the $\ia$ and $\ib$ indices, $\phi_{\ia\ib}$ is symmetric
and $\phi_{34}=\phi_{12}$, $\phi_{24}=\phi_{13}$ and
$\phi_{23}=\phi_{14}$, while
$\phi_{11}=\phi_{22}=\phi_{33}=\phi_{44}=0$.  The $X_\ia$ are then
given by
\begin{equation}
   X_\ia = \exp(-\vec{a}_\ia\cdot\vec{\phi}/2) ,
\label{eq:phis}
\end{equation}
where
\begin{equation}
\begin{aligned}
   \vec{a}_1 &= (1,1,1) , & \quad  \vec{a}_2 &= (1,-1,-1) , \\
   \vec{a}_3 &= (-1,1,-1) , & \quad  \vec{a}_4 &= (-1,-1,1) .
\end{aligned}
\label{eq:ai}
\end{equation}

As discussed in~\cite{duff10}, this is the bosonic action of $d=4$
$N=2$ gauged $U(1)^4$ supergravity with the three axions set to
zero. Thus the reduction ansatz can be used to embed $d=4$ solutions
with vanishing axions into solutions of eleven-dimensional
supergravity. The same bosonic action can also be obtained by
truncating $N=8$ gauged $SO(8)$ supergravity as
in~\cite{duffliu}. (This is where the peculiar index structure of
$\phi_{\ia\ib}$ comes from: these fields really parametrise a
self-dual four-form under $SO(8)$.) The corresponding $N=8$ fermionic
supersymmetry transformations can be written as follows. First, note
that the $N=8$ fermions consist of the gravitini $\psi_\mu^I$ and the
spin-half fields $\chi^{IJK}$ where $I$, $J$ and $K$ are $SU(8)$
indices. Given the ansatz for the scalar and vector fields, the index
$I$ is equivalent to the pair $(\ia,i)$ where $\ia=1,\dots,4$ as above
and $i=1,2$. With this notation the variations of the gravitini are
given by~\cite{duffliu}
\begin{equation}
\begin{aligned}
   \delta\psi_\mu{}^{\ia i} &=
       \nabla_\mu\epsilon^{\ia i}
       - 2e \sum_\ib \Omega_{\ia\ib} A_{\ib,\mu} \epsilon^{ij}
           \epsilon^{\ib j}
       + \frac{e}{4\sqrt{2}} \sum_\ib \me^{-\vec{a}_\ib\cdot \vec{\phi}/2}
           \gamma_\mu\epsilon^{\ia i}
   \\ & \qquad
       + \frac{1}{2\sqrt{2}} \sum_\ib \Omega_{\ia\ib}
           \me^{\vec{a}_\ib\cdot\vec{\phi}/2} F_{\ib,\nu\lambda}
           \gamma^{\nu\lambda}\gamma_\mu
           \epsilon^{ij}\epsilon^{\ia j} ,
\end{aligned}
\label{eq:gravsusy}
\end{equation}
where $\Omega_{\ia\ib}$ is the matrix
\begin{equation}
   \Omega = \frac{1}{2}\left( \begin{array}{rrrr}
       1 & 1 & 1 & 1   \\  1 & 1 & -1 & -1 \\
       1 & -1 & 1 & -1 \\  1 & -1 & -1 & 1
       \end{array} \right) .
\label{eq:Omegadef}
\end{equation}
For the spin-1/2 fermions, one finds
$\delta\chi^{\ia i\;\ib j\;\ic k} =
\delta\chi^{\ia\;\ic k}\delta^{\ia\ib}\epsilon^{ij} +
\delta\chi^{\ib\;\ia i}\delta^{\ib\ic}\epsilon^{jk} +
\delta\chi^{\ic\;\ib j}\delta^{\ic\ia}\epsilon^{ki}$ with~\cite{duffliu}
\begin{equation}
\begin{aligned}
   \delta\chi^{\ia\;\ib i} &=
       - \frac{1}{\sqrt{2}} \gamma^\mu \partial_\mu\phi_{\ia\ib}
           \epsilon^{ij}\epsilon^{\ib j}
       - e \sum_{\ic\id} \Sigma_{\ia\ib\ic} \Omega_{\ic\id}
           \me^{-\vec{a_\id}\cdot\vec{\phi}/2}
           \epsilon^{ij}\epsilon^{\ib j}
   \\ & \qquad
       + \sum_\id \Omega_{\ia\id} \me^{\vec{a}_\id\cdot\vec{\phi}/2}
           F_{\id\,\mu\nu}\gamma^{\mu\nu} \epsilon^{\ib i} .
\end{aligned}
\label{eq:chisusy}
\end{equation}
The tensor $\Sigma_{\ia\ib\ic}$ selects a particular $\ic$ depending on
$\ia$ and $\ib$, and is defined by
\begin{equation}
\Sigma_{\ia\ib\ic} = \begin{cases}
    |\epsilon_{\ia\ib\ic}| & \text{for } \ia,\ib,\ic \ne 1 , \\
    \delta_{\ib\ic} & \text{for } \ia = 1 , \\
    \delta_{\ia\ic} & \text{for } \ib = 1 , \\
    0 & \text{otherwise} .
    \end{cases}
\end{equation}
%

%%%%%%%%%%%%%%%%%%%%%%%%%%%%%%%%%%%%%%%%

\section{BPS Equations}

Our strategy for constructing supersymmetric $D=11$ solutions
describing wrapped membranes is to first construct four-dimensional BPS
solutions of \eqref{eq:S} such that the supersymmetry
variations~\eqref{eq:gravsusy} and~\eqref{eq:chisusy} vanish. These
are then uplifted to $D=11$ using ansatz~\eqref{eq:g11}
and~\eqref{eq:G4}. For orientation note that the vacuum $AdS_4$
solution
\begin{equation}
   \dif s^2 = \frac{1}{2e^2 r^2} \left( - \dif t^2 + \dif x^2 + \dif y^2 + \dif r^2 \right),
\end{equation}
with the gauge fields and the scalars set to zero, uplifts to
$AdS_4\times S^7$, the supergravity dual of the SCFT for flat
unwrapped membranes. 

To describe wrapped membranes, we take the ansatz for the
four-dimensional metric to be 
\begin{equation}
   \dif s^2 = \me^{2f} \left( - \dif t^2 + \dif r^2 \right) 
      + \me^{2g} \dif s^2(\Sigma) ,
\label{eq:metric}
\end{equation}
where $\dif s^2(\Sigma)$ is the metric on the two-cycle $\Sigma$.
The ansatz for the gauge fields is determined as follows.
First recall that the normal bundle to a holomorphic two-cycle
in a Calabi--Yau $(n+1)$-fold is $U(n)\subset SO(2n)$. Moreover, for
the Calabi--Yau manifold to have only $SU(n+1)$ holonomy, the $U(1)$
spin connection of $\Sigma$ must be identified with
the diagonal $U(1)$ factor in  $U(n)$.
When a membrane probe wraps the two-cycle, the
$SO(8)$ R-symmetry of the normal directions of an unwrapped
membrane in flat space, is naturally split into
$SO(2n)\times SO(8-2n)$, reflecting the
split of the directions normal to the two-cycle
within the Calabi--Yau manifold and the rest. The structure of the normal
bundle of the Calabi--Yau manifold then automatically requires an
identification of the $U(1)$ spin connection on the cycle with the
corresponding diagonal $U(1)\subset U(n)\subset SO(2n)$ part of the
R-symmetry: this is what is meant by ``twisting''. The upshot of these
observations is that in the supergravity ansatz the only non-vanishing
$SO(8)$ gauge fields should be lie in $U(n)\subset SO(2n)$ and we must
identify the $U(1)$ spin connection of $\Sigma$ in \eqref{eq:metric} with
the diagonal $U(1)\subset U(n)$. Now the above supergravity truncation
keeps only Abelian gauge fields $U(1)^4\subset SO(8)$. We thus want to
identify each of the $n$ $U(1)$ gauge fields in $SO(2n)$ with the
$U(1)$ spin connection, and set the remaining $U(1)$ gauge fields to
zero. Note that the restriction to Abelian gauged supergravity means,
for $n\ge2$, that we are considering twistings corresponding to
non-generic Calabi--Yau $(n+1)$-folds for which the normal bundle to
the two-cycle is restricted to be only $U(1)$ rather than the full
$U(n)$. A familiar Calabi--Yau three-fold example is provided by the
resolved conifold.  

The supergravity ansatz is completed by specifying scalar fields
consistent with the $SO(2n)\times SO(8-2n)$ split. In each case
this is achieved by writing $\vec \phi$ in terms of
a single scalar field $\phi$.

Requiring the fermionic variations~\eqref{eq:gravsusy}
and~\eqref{eq:chisusy} to vanish with this ansatz
leads to a set of BPS first-order
differential equations for the metric functions $f$ and $g$ and the
scalar field $\phi$. These shall be presented in the subsequent
subsections. In all cases, we find that the Killing spinors
$\epsilon^{\ia i}$ satisfy
\begin{equation}
   \label{eq:projections}
   \gamma^3 \epsilon^{\ia i} = \epsilon^{\ia i} ,
   \qquad
   \epsilon^{\ia i} = \me^{f/2} \epsilon_0^{\ia i} ,
\end{equation}
where $\epsilon_0^{\ia i}$ is a constant spinor and $\gamma^3$ points
in the radial direction and is defined using the orthonormal frame 
\begin{equation}
   \label{eq:frame}
   e^m = (\me^f \dif t, \me^g \bar e^1, \me^g \bar e^2, \me^f \dif r) ,
\end{equation}
with $(\bar e^1, \bar e^2)$ giving an orthonormal frame for $\Sigma$. 
The first condition breaks half of the supersymmetry. In each case,
there are then different additional restrictions, breaking more
supersymmetry. The second condition comes from the $r$ component of the
gravitino variation~\eqref{eq:gravsusy}. Otherwise the Killing spinors
are independent of coordinates. As in previous studies
\cite{malnun,malnuntwo,agk,no,gkw}, it is
necessary that the metric on the 2-cycle is Einstein which means that
it has constant curvature. The cycle is either a sphere $S^2$,
hyperbolic space $H^2$ or flat.
The Ricci tensor is given by
\begin{equation}
   R(\Sigma)_{ab} = \lk g_{ab} ,
\label{eq:kdef}
\end{equation}
where $a$ and $b$ are indices on $\Sigma$ and the volume of the cycle
is normalised so that $\lk=1$ for $\Sigma=S^2$, $\lk=0$ for
$\Sigma=\bbR^2$ and $\lk=-1$ for $\Sigma=H^2$. Since the $D=4$ Killing
spinors are independent of the coordinates of the cycle, we can also
take quotients of these spaces while preserving supersymmetry.
In particular $\Sigma$ can be a compact Riemann surface of any genus.

Before discussing the different cases let us comment on
the flat case, $l=0$. The fact that we are considering
gauge fields corresponding to Calabi--Yau manifolds with non-generic normal
bundles means that if $\Sigma$ is flat then the whole normal bundle is
in fact trivial. Thus the Calabi--Yau manifold is locally just flat
$\bbC^{d+1}$, and in all cases we are simply considering the embedding
of a flat M2-brane in flat space. The corresponding supergravity
solution is then very well known
and can be written in familiar form in terms of a harmonic
function. For this case solutions to our BPS equations
simply correspond to a choice of harmonic function preserving
$SO(2n)\times SO(8-2n)$ symmetry in the eight transverse dimensions.
In fact it is straightforward to solve the $l=0$ BPS equations exactly
by introducing isotropic coordinates as in the Calabi--Yau five-fold case
discussed below. As a consequence we shall not dwell on the $l=0$ case 
in the sequel. Note that these solutions are the direct analogues of
the D3-brane solutions given in~\cite{FGPW}.

\subsection{Calabi--Yau two-fold}

For this case there is a natural split of $SO(8)$ into $SO(2)\times SO(6)$.
The non-vanishing gauge fields are in the $SO(2)$ gauge group and so we have
\begin{equation}
   F_1 = -\frac{\lk}{2e}\Vol(\Sigma), \qquad
   F_2 = F_3 = F_4 = 0 ,
\end{equation}
where $\Vol(\Sigma)$ is the volume two-form on the two-cycle. The
scalar fields are given by $\vec{\phi}=(\phi,\phi,\phi)$ 
so that $X_1=\me^{-{3\phi/2}}$, $X_2=X_3=X_4=\me^{\phi/2}$.

For a two-cycle in a Calabi--Yau two-fold, we expect to preserve
eight supercharges. Demanding that the supersymmetry variations
are zero, we find that, in addition to \eqref{eq:projections},
the Killing spinor $\epsilon^{\ia i}$ must satisfy
\begin{equation}
   \gamma^{12}\epsilon^{\ia i} = \epsilon^{ij} \epsilon^{\ia j}
   \quad \text{for $\ia=1,\dots,4$} ,
\label{eq:spinorCY2}
\end{equation}
where $1$ and $2$ are tangent space indices in the $\Sigma$ directions, 
as defined by the orthonormal frame~\eqref{eq:frame}. Together these
conditions do indeed give eight independent Killing spinors. The
variations then vanish provided we satisfy the BPS equations
\begin{equation}
\begin{aligned}
   \me^{-f}f' &= -{e\over 2{\sqrt 2}}(\me^{-3\phi/2}+3\me^{\phi/2})
     +{\lk \over 2{\sqrt 2 }e}{\me^{3 \phi/2-2g}} , \\
   \me^{-f}g' &= -{e\over 2{\sqrt 2}}(\me^{-3\phi/2}+3\me^{\phi/2})
     -{\lk \over 2{\sqrt 2 }e}{\me^{3 \phi/2-2g}} , \\
   \me^{-f}\phi'&= -{e\over {\sqrt 2}}(\me^{-3\phi/2}-\me^{\phi/2})
     +{\lk \over {\sqrt 2 }e}{\me^{3 \phi/2-2g}} .
\end{aligned}
\label{eq:BPS2in4}
\end{equation}

These equations can be partially integrated to give 
\begin{equation}
   \me^{2g+\phi} = C \left( 2\me^{2g-\phi}+{\lk\over e^2} \right)^{1/2}
      + \me^{2g-\phi}+{\lk\over e^2} ,
\label{eq:Fxsol}
\end{equation}
where $C$ is a constant. It is also straightforward to determine the
full asymptotic behaviour of the solutions. In different limits, different
terms in the BPS equations~\eqref{eq:BPS2in4} dominate. For
example, for large $\me^{2g}$, the final terms proportional to $\lk$ are
small and the leading behaviour, valid at small $r$, is given by 
\begin{equation}
   \dif s^2 \approx \frac{1}{2e^2r^2} \left(
           - \dif t^2 + \dif r^2 + \dif s^2(\Sigma) \right) .
\label{ads4}
\end{equation}
with $\phi\approx cr+\left(\lk-\frac{1}{2}c^2\right)r^2$, for an
arbitrary constant $c$. This is almost $AdS_4$ except the metric on
$\bbR^{1,2}$ has been replaced with one on $\bbR\times \Sigma$. This
limit specifies the UV behaviour of the membrane wrapped on $\Sigma$,
and is universal in the sense that, as we will see, it is present
independent of the dimension of the Calabi--Yau manifold and of the
curvature of $\Sigma$. The leading and sub-leading terms in $\phi$
are interpreted as either the insertion of a boundary operator in the
UV theory, due to the curvature of the two-cycle, or the expectation
value of this operator~\cite{kw}.  

\begin{figure}[htbp]
   \begin{center}
      \epsfig{file=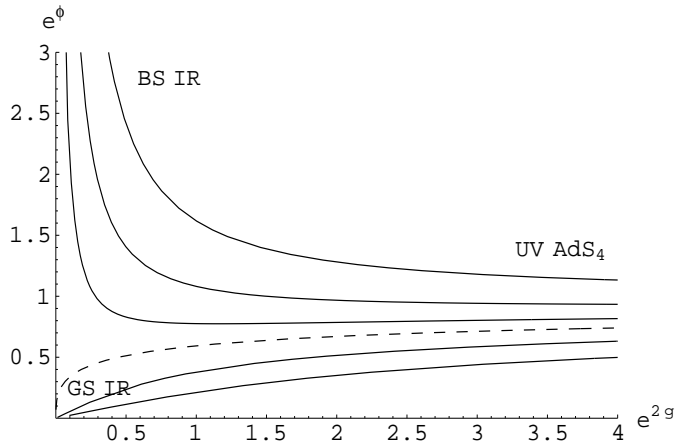,width=9cm,height=6cm}
      \caption{Behaviour of the flows for an $S^2$ cycle in a
        Calabi--Yau two-fold \label{fig:4S2}}
   \end{center}
\end{figure}

\begin{figure}[htbp]
   \begin{center}
      \epsfig{file=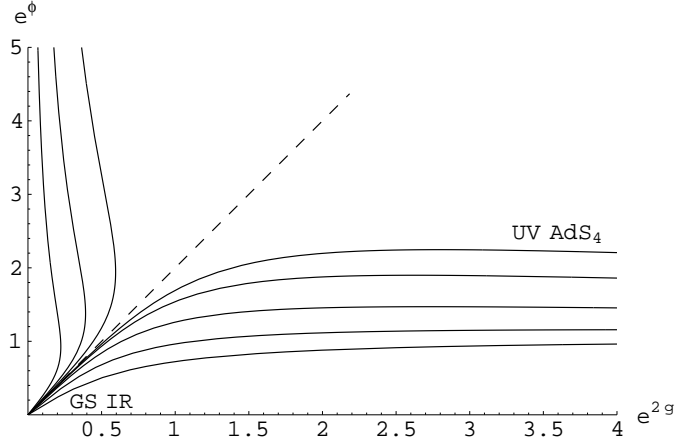,width=9cm,height=6cm}
      \caption{Behaviour of the flows for an $H^2$ cycle in a
        Calabi--Yau two-fold \label{fig:4H2}}
   \end{center}
\end{figure}

The behaviour of the solutions are illustrated in figures~\ref{fig:4S2}
and~\ref{fig:4H2} for the case of $\Sigma=S^2$ and $\Sigma=H^2$,
respectively, for different values of $C$ in~\eqref{eq:Fxsol}. These
are interpreted as describing the flows from the UV region to gravity
duals of the IR physics. In each case, we have noted whether the resulting
singularities encountered in the IR are of good or bad type according
to the criteria of \cite{malnun}. Recall that for ``good''
singularities the time component of the uplifted eleven-dimensional
metric $g^{(11)}_{00}$ goes to zero for the strong form of ``good''
and to a constant for the weak form. For a ``bad'' singularity
$g^{(11)}_{00}$ is unbounded. The physical idea behind the criteria
of good singularities is that one expects that, as one goes to the IR,
fixed proper energies should correspond to smaller or non-increasing
energies in the dual field theory. The good singularities should 
correspond to different physical branches of the dual IR quantum
mechanics theory and the bad singularities to non-physical
solutions. We shall return briefly to this in the final section. 

Explicitly, for both $\Sigma=S^2$ and $\Sigma=H^2$ there
is a bad singularity when $\me^{2g}\to 0$ and $\me^\phi\to\infty$, where
the BPS equations are dominated by the final terms proportional to
$\lk$. (This limit is also universal, present whatever the
dimension of the Calabi--Yau manifold and the value of $\lk$. However,
in all cases, only for $\Sigma=S^2$ is there a flow from the UV
$AdS_4$-type region to this singularity.) For either value of $\lk$
there is also a good singularity when both $\me^{2g}$ and $\me^\phi$ tend
to zero, where one can neglect the $\me^{\phi/2}$ terms in the BPS
equations. For $\lk=1$, the asymptotic solution has the form
\begin{equation}
\begin{aligned}
   \dif s^2 &= - \left( \frac{z^2-z_0^2}{4e^{-1}z} \right)^{1/2} \dif t^2
       + \left( \frac{4e^{-1}z}{z^2-z_0^2} \right)^{1/2} \left[
            \dif z^2 + \left(z^2-z_0^2\right)\dif s^2(\Sigma) \right] , \\
   \me^\phi &= \left( \frac{z^2-z_0^2}{e^{-1}z} \right)^{1/2} ,
\end{aligned}
\label{eq:4S2good}
\end{equation}
where $z_0$ is a constant, such that $z_0\geq 0$ and the solution is
valid only in leading order with $z\to z_0$ from above. For $\lk=-1$
the solution has the same form but with $z^2-z_0^2$ replaced with
$z_0^2-z^2$, with $z_0>0$ and $z\to z_0$ from below. 
In terms of the parameter $C$ in~\eqref{eq:Fxsol}, for $\Sigma=S^2$,
solutions with $C\leq -\sqrt{\lk/e^2}$ flow to the good singularity,
while those with $C>-\sqrt{\lk/e^2}$ flow to the bad singularity. For
$\Sigma=H^2$, whatever the value of $C$ all solutions flow to the good
singularity. 

\subsection{Calabi--Yau three-fold}

For this case, there is a natural split of $SO(8)$ into
$SO(4)\times SO(4)$. The diagonal $U(1)$ gauge fields
of $U(2)\subset SO(4)$ are non-vanishing and so we take
\begin{equation}
   F_1=F_2=-{\lk\over 4e}\Vol(\Sigma) ,
   \qquad
   F_3=F_4=0 ,
\end{equation}
where, as before, $\Vol(\Sigma)$ is the volume two-form on the
two-cycle. The scalar fields are given by
$\vec{\phi}=(\phi,0,0)$, so that $X_1=X_2=\me^{-\phi/2}$ and
$X_3=X_4=\me^{\phi/2}$. We expect a theory with four supercharges, and
indeed find that setting the fermionic variations to zero requires, in
addition to \eqref{eq:projections}, the projections
\begin{equation}
\begin{aligned}
   \gamma^{12} \epsilon^{\ia i} &= \epsilon^{ij} \epsilon^{\ia i}
       \quad && \text{for $\ia=1,2$} , \\
   \epsilon^{\ia i} &= 0 \quad && \text{for $\ia=3,4$} .
\end{aligned}
\label{eq:spinorCY3}
\end{equation}
The variations then vanish provided we satisfy the BPS equations 
\begin{equation}
\begin{aligned}
   \me^{-f}f' &= -{e\over {\sqrt 2}}(\me^{-\phi/2}+\me^{\phi/2})
       +{\lk \over 2{\sqrt 2 }e}{\me^{\phi/2-2g}} , \\
   \me^{-f}g' &= -{e\over {\sqrt 2}}(\me^{-\phi/2}+\me^{\phi/2})
       -{\lk \over 2{\sqrt 2 }e}{\me^{\phi/2-2g}} , \\
   \me^{-f}\phi' &= -{\sqrt 2}e(\me^{-\phi/2}-\me^{\phi/2})
       +{\lk \over {\sqrt 2 }e}{\me^{\phi/2-2g}} .
\end{aligned}
\label{eq:BPS6}
\end{equation}

To analyse these equations, one can introduce the variables 
$x=\me^{2g-\phi}$, $F=\me^{g+\phi/2}$ which then satisfy
\begin{equation}
{\dif F\over \dif x}={F\over 2F{\sqrt x}+\lk/e^2} .
\end{equation}
It is possible to solve this equation exactly, but the resulting form
is not very illuminating. It is anyway again straightforward
to determine the asymptotic behaviour of the solutions.
\begin{figure}[htbp]
   \begin{center}
      \epsfig{file=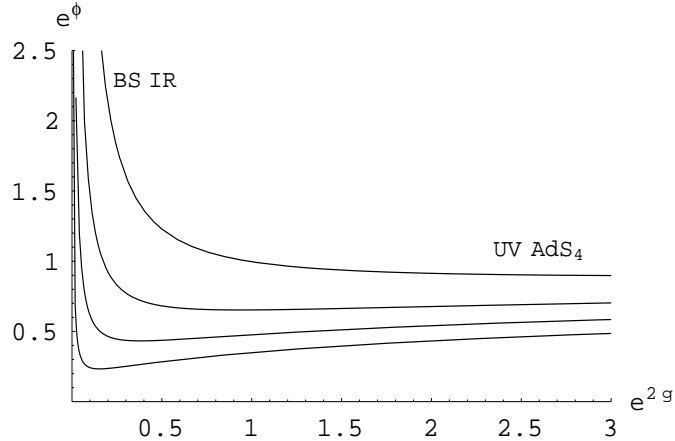,width=9cm,height=6cm}
      \caption{Behaviour of the flows for an $S^2$ cycle in a
        Calabi--Yau three-fold\label{fig:6S2}}
   \end{center}
\end{figure}

\begin{figure}[htbp]
   \begin{center}
      \epsfig{file=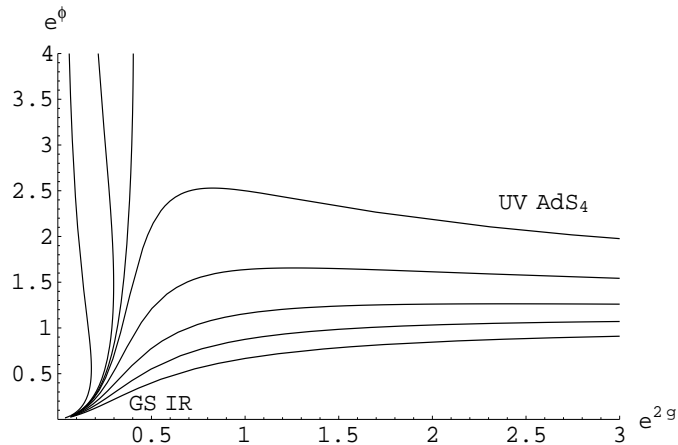,width=9cm,height=6cm}
      \caption{Behaviour of the flows for an $H^2$ cycle in a
        Calabi--Yau three-fold \label{fig:6H2}}
   \end{center}
\end{figure}

For large $\me^{2g}$ we have the $AdS_4$-type region \eqref{ads4}
describing the UV physics, now with $\phi\approx cr+\lk r^2$, where $c$
is an arbitrary constant. We have illustrated the 
flows to the IR and the resulting types of singularities in
figures~\ref{fig:6S2} and~\ref{fig:6H2}. As before, there are bad
singularities in the regions of large $\me^\phi$ and small $\me^{2g}$
where the BPS equations~\eqref{eq:BPS6} are dominated by the terms
proportional to $\lk$. There is a good singularity only for
$\Sigma=H^2$, in the region of small $\me^\phi$ and $\me^{2g}$ where one
can neglect the $\me^{\phi/2}$ terms in~\eqref{eq:BPS6}. The asymptotic
solution is given by 
\begin{equation}
\begin{aligned}
   \dif s^2 &=  \left(\frac{r_0}{r}\right)^{1/2} \me^{-2er} \left(
      - \dif t^2 + \dif r^2 + e^{-1}r \; \dif s^2(\Sigma) \right) , \\
   \me^\phi &= \left(\frac{r_0}{4r}\right)^{1/2} \me^{-2er} ,
\end{aligned}
\label{eq:6H2good}
\end{equation}
where $r_0$ is a constant.

\subsection{Calabi--Yau four-fold}

In this case the $SO(8)$  splits into $SO(6)\times SO(2)$. The only
non-zero gauge fields are in the diagonal $U(1)$ in $U(3)\subset
SO(6)$ and so we have 
\begin{equation}
   F_1=F_2=F_3=-{\lk\over 6e}\Vol(\Sigma),\qquad F_4=0 ,
\end{equation}
with $\Vol(\Sigma)$ is the volume two-form on $\Sigma$.
The scalar fields are given by $\vec{\phi}=(\phi,\phi,-\phi)$, so that
$X_1=X_2=X_3=\me^{-\phi/2}$ and $X_4=\me^{3\phi/2}$. Requiring the
fermionic variations to vanish implies in addition to
\eqref{eq:projections} the projections 
\begin{equation}
\begin{aligned}
   \gamma^{12} \epsilon^{1 i} &= \epsilon^{ij} \epsilon^{1 i} , \\
   \epsilon^{\ia i} &= 0 \quad \text{for $\ia=2,3,4$} ,
\end{aligned}
\label{eq:spinorCY4}
\end{equation}
which gives a theory with two supercharges, as expected. In addition,
we find the BPS equations
\begin{equation}
\begin{aligned}
   \me^{-f}f' &= -{e\over 2{\sqrt 2}}(3\me^{-\phi/2}+\me^{3\phi/2})
       + {\lk \over 2{\sqrt 2 }e}{\me^{\phi/2-2g}} , \\
   \me^{-f}g' &= -{e\over 2{\sqrt 2}}(3\me^{-\phi/2}+\me^{3\phi/2})
       - {\lk \over 2{\sqrt 2 }e}{\me^{\phi/2-2g}} , \\
   \me^{-f}\phi' &= -{e\over {\sqrt 2}}(\me^{-\phi/2}-\me^{3\phi/2})
       + {\lk \over 3{\sqrt 2 }e}{\me^{\phi/2-2g}} .
\end{aligned}
\label{eq:BPS8}
\end{equation}

\begin{figure}[htbp]
   \begin{center}
      \epsfig{file=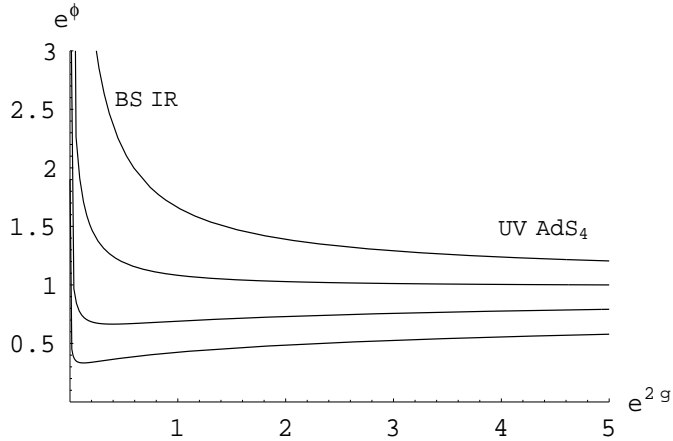,width=9cm,height=6cm}
      \caption{Behaviour of the flows for an $S^2$ cycle in a
        Calabi--Yau four-fold \label{fig:8S2}}
   \end{center}
\end{figure}

As before it is straightforward to analyse the asymptotic
behaviour. For large $\me^{2g}$ there is the UV region $AdS_4$-type
region~\eqref{ads4} now with $\phi\approx
cr+\left(\frac{1}{3}\lk+\frac{1}{2}c^2\right)r^2$ where $c$ is an
arbitrary constant. For $\Sigma=S^2$ there is only one other
asymptotic region at  small $\me^{2g}$ and large $\me^\phi$, where
there is a bad singularity and the BPS equations are dominated by the
terms proportional to $\lk$. The general flows are shown in
figure~\ref{fig:8S2}.

\begin{figure}[htbp]
   \begin{center}
      \epsfig{file=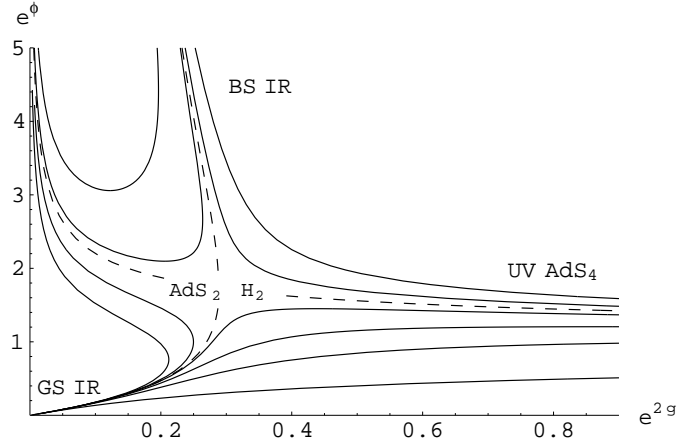,width=9cm,height=6cm}
      \caption{Behaviour of the flows for an $H^2$ cycle in a
        Calabi--Yau four-fold \label{fig:8H2}}
   \end{center}
\end{figure}

For the $\Sigma=H^2$ the situation is considerably more complicated,
as shown in figure~\ref{fig:8H2}. Aside from the UV $AdS_4$-type
region, there is also, for the first time, an $AdS_2\times H^2$ fixed
point. Explicitly, there is an exact solution of~\eqref{eq:BPS8} given by 
\begin{equation}
\begin{aligned}
   \dif s^2 &= \frac{1}{6\sqrt{3}e^2r^2} \left( - \dif t^2 + \dif r^2 \right)
      + \frac{1}{2\sqrt{3}e^2} \dif s^2(H^2) , \\
   \me^\phi &= \sqrt{3} .
\end{aligned}
\label{eq:adsfp}
\end{equation}
This provides an example of ``flow across dimensions'' from a
three-dimensional UV theory to a superconformal quantum mechanics IR
fixed point.  

In addition, there are two regions of bad singularities both at small
$\me^{2g}$ and large $\me^\phi$. One is the universal bad singularity
where the terms proportional to $\lk$ in~\eqref{eq:BPS8} dominate. The
other corresponds to a region where $\lk$ and $\me^{-\phi/2}$ terms are
proportional and dominate. There is also a region at small $\me^{2g}$ and
$\me^\phi$ where one can neglect the $\me^{3\phi/2}$ terms and which gives
a good singularity. The asymptotic solution is
\begin{equation}
\begin{aligned}
   \dif s^2 &= \frac{r_0^3}{r^3} \left( - \dif t^2 + \dif r^2
           + \tfrac{3}{4}r^2 \dif s^2(\Sigma) \right) , \\
   \me^\phi &= \frac{9e^2r_0^3}{8r} ,
\end{aligned}
\label{eq:8H2good}
\end{equation}
where $r_0$ is a constant.

\subsection{Calabi--Yau five-fold}

For this case the diagonal $U(1)$ of $U(4)\subset SO(8)$
is the only non-vanishing gauge field giving rise to
\begin{equation}
   F_1 = F_2 = F_3 = F_4 = -{\lk\over 8e}\Vol(\Sigma) ,
\end{equation}
where $\Vol(\Sigma)$ is the volume two-form on $\Sigma$.
The scalar fields are all zero so that $X_i=1$.
The projections on the Killing spinors are exactly the same as in the
Calabi--Yau four-fold case~\eqref{eq:projections},~\eqref{eq:spinorCY4},
leading to preservation of two supercharges. 
The BPS equations then read
\begin{equation}
\begin{aligned}
   \me^{-f}f' &= -{\sqrt 2}e + {\lk \over 2{\sqrt 2 }e}{\me^{-2g}} , \\
   \me^{-f}g' &= -{\sqrt 2}e - {l \over 2{\sqrt 2 }e}{\me^{-2g}} ,
\end{aligned}
\end{equation}
We can find the general solution by first introducing a new
radial variable defined by
\begin{equation}
   {\dif\rho\over \dif r}=\me^{2f} .
\end{equation}
We then obtain, after absorbing an integration constant into the
definition of the coordinate $t$,
\begin{equation}
   \dif s^2=- \left(\sqrt{2}e\rho
              + \frac{l}{2\sqrt{2}e\rho}\right)^2 \dif t^2
            + \left( \sqrt{2}e\rho
              + \frac{l}{2\sqrt{2}e\rho}\right)^{-2} \dif\rho^2
            + \rho^2 \dif s^2(\Sigma) .
\label{full}
\end{equation}
When $\lk=1$ we obtain a bad IR singularity in the IR.
When $\lk=-1$ the solution interpolates from the $AdS_4$ type
region to the superconformal $AdS_2\times H^2$ fixed point in the IR
specified by 
\begin{equation}
\begin{aligned}
   \dif s^2 &= \frac{1}{8e^2r^2} \left( - \dif t^2 + \dif r^2 \right)
      + \frac{1}{4e^2} \dif s^2(H^2) .
\end{aligned}
\end{equation}
This IR fixed point is the gravity dual of a superconformal quantum mechanics.

We note that when $\lk=-1$ the full solution \eqref{full} is the
supersymmetric magnetically charged ``topological'' AdS black hole
discussed in \cite{caldklemm}. The term topological refers to the
unusual feature that black holes in AdS space can admit spatial
sections that are flat or have constant negative curvature. Here we
see that this solution can be interpreted, after being uplifted to
$D=11$, as the gravity dual corresponding to wrapped membranes in a
Calabi--Yau five-fold. It was also observed in \cite{caldklemm} that
the four-dimensional solution with $\lk=-1$ can be generalised to
include rotation while maintaining supersymmetry. In our conventions
it is given by 
\begin{equation}
\begin{aligned}
    \dif s^2 = -& \frac{\Delta_r}{\Xi^2\rho^2}
            \left[\dif t + a\sinh^2\theta \dif\phi\right]^2
       + \frac{\rho^2}{\Delta_r} \dif r^2
       + \frac{\rho^2}{\Delta_\theta} \dif\theta^2 \\
       &+ \frac{\Delta_\theta\sinh^2\theta}{\Xi^2\rho^2}
            \left[a\dif t - (r^2+a^2) \dif\phi\right]^2 ,
\end{aligned}
\end{equation}
with
\begin{equation}
\begin{aligned}
   \rho^2 &= r^2 + a^2\cosh^2\theta , &\qquad
   \Xi &= 1 + 2e^2a^2 , \\
   \Delta_r &= r^2 \left[\sqrt{2}er
         - \frac{1}{2\sqrt{2}er} \left(1-2e^2a^2\right)\right]^2, &\qquad
   \Delta_\theta &= 1 + 2e^2a^2\cosh^2\theta ,
\end{aligned}
\end{equation}
and
\begin{equation}
   A_1 = - \frac{(1+2e^2a^2)\cosh\theta}{8e\Xi\rho^2}
              \left[ a \dif t - (r^2+a^2) \dif\phi \right] .
\end{equation}
An interesting feature of this solution is that it is regular provided
that the rotation parameter $a$ satisfies $a<\sqrt{2}e$.
We can use this four-dimensional solution to obtain a new
$D=11$ supersymmetric solution by uplifting using the
ansatz~\eqref{eq:g11} and~\eqref{eq:G4}. This $D=11$ solution can be
interpreted as the gravity  dual corresponding to supersymmetric waves
on wrapped membranes. The bound on the angular momentum parameter is
reminiscent of the ``stringy exclusion principle'' observed in
\cite{malstrom} and it would be interesting to explore this in more
detail.

%%%%%%%%%%%%%%%%%%%%%%%%%%%%%%%%%%%%%%%%

\section{Generalised Calibrations}

Our $D=11$ solutions correspond to the near-horizon geometry of
membranes wrapping holomorphic two-cycles in a Calabi--Yau
$(n+1)$-fold. Being supersymmetric we expect that a probe membrane will
be static and supersymmetric if it wraps the same two-cycle,
or more precisely a holomorphic cycle in the same homology class. In
the language of~\cite{gpt} this means that we expect that the $D=11$
solutions admit generalised K\"{a}hler two-form calibrations
$\Omega$. A static, supersymmetric probe brane then minimises the
pull-back of $\Omega$ integrated over the two-cycle. The calibration
$\Omega$ can be constructed from $D=11$ Killing spinors $\epsilon$ via
$\Omega_{MN}=\bar\epsilon\Gamma_{MN}\epsilon$.  

Let us give an explicit expression for $\Omega$ for our solutions and
show that it is indeed a calibration. First it is useful to introduce
a slightly non-obvious $D=11$ orthonormal frame 
\begin{equation}
\begin{gathered}
   e^0 = \Delta^{1/3}\me^f \dif t , \\
   e^1 = \Delta^{1/3}\me^g \bar{e}^1 , \\
   e^2 = \Delta^{1/3}\me^g \bar{e}^2 , \\
\begin{aligned}
   e^{\rho_\ia} &= \Delta^{-1/6}\left[
       \me^{f}X^{1/2}_\ia\mu_\ia \dif r
       - \sqrt{2}e^{-1}X^{-1/2}_\ia \dif\mu_\ia \right] ,\\
   e^{\phi_\ia} &=
       \sqrt{2}e^{-1}\Delta^{-1/6} X^{-1/2}_\ia \mu_\ia
           \left(\dif\phi_\ia+2e A_\ia\right) ,
\end{aligned}
\end{gathered}
\label{eq:ortho}
\end{equation}
where, as before,  $\bar{e}^1$ and $\bar{e}^2$ define an orthonormal
frame for the two-cycle. We then have, in all cases, 
\begin{equation}
   \Omega = \Delta^{1/3}\me^f \left[
      e^1\wedge e^2 + \sum_\ia e^{\rho_\ia}\wedge e^{\phi_\ia} \right].
\end{equation}
For the directions $\ia$ with vanishing gauge fields we find
$\dif[\Delta^{1/3}\me^f\; e^{\rho_\ia}\wedge e^{\phi_\ia}]=0$.
Using this, and the relevant BPS equations, we can show that
\begin{equation}
   \dif\Omega= i_k G.
\label{gencal}
\end{equation}
where $k=\partial/\partial t$. This is one of the conditions satisfied
by the generalised calibration (and in fact follows from the spinorial
construction). In addition, we note that $\Omega$ naturally defines an
almost complex structure $J$ on the ten-dimensional spatial part of
our solution. In the orthonormal frame, $J$ simply pairs $e^1$ with
$e^2$ and $e^{\rho_\alpha}$ with $e^{\phi_\alpha}$. Formally it can be
defined by raising one index of $\Omega$ using the rescaled $D=10$
spatial metric  
\begin{equation}
   \dif\tilde s^2=\Delta^{1/3}\me^f \left[
      e^1e^1 + e^2e^2 + \sum_\ia e^{\rho_\ia} e^{\rho_\ia}
      + \sum_\ia e^{\phi_\ia} e^{\phi_\ia} \right] .
\end{equation}
We have checked that it is in fact integrable (without using the BPS
equations) and so the ten-dimensional spatial part of our solutions in
fact describes a complex manifold. This and equation~\eqref{gencal}
establish that $\Omega$ is indeed a generalised
calibration~\cite{gpt}. Note, in addition, as is easy to see in the
orthonormal frame, the spatial part of our metric is Hermitian with
respect to this complex structure.  

An alternative approach to finding solutions corresponding to  
membranes wrapping holomorphic curves in Calabi--Yau $(n+1)$-folds was
discussed in~\cite{cekt}. Building on the work of~\cite{FS1} and
exploiting the existence of a generalised calibration an ansatz for
the solutions was presented. The BPS equations were derived but no
solutions were given. The ansatz has the form
\begin{equation}
\begin{aligned}
   \dif s^2 &= - H^{-2n/3}\dif t^2 
      + H^{(n-3)/3} g_{AB} \dif y^A \dif y^B
      + H^{n/3} \delta_{IJ} \dif x^I \dif x^J ,  \\
   A_{(3)} &= \pm H^{-1} \dif t \wedge \omega ,
\end{aligned}
\label{eq:cekt}
\end{equation}
where $y^A$ with $A=1,\dots,2n+2$ are real coordinates on a complex
$(n+1)$-fold with a Hermitian metric $g_{AB}$ (the analogue of the
original Calabi--Yau manifold), and $x^I$ with $I=1,\dots,8-2n$ denote
the remaining transverse directions. The two-form $\omega$ is related to
the Hermitian metric $g_{AB}$ by $\omega_{AB}={J^C}_Ag_{CB}$ where $J$
is the complex structure on the $(n+1)$-fold. Both $g_{AB}$ and
$\omega_{AB}$ are functions of $y^A$ and $x^I$ as is
$H$. Supersymmetry then puts various constraints on $g_{AB}$ and $H$,
for instance implying that for fixed $x^I$ the metric $g_{AB}$ is
K\"ahler.  

To connect this work to our solutions, we would like show that they 
can be written in the form~\eqref{eq:cekt}. Starting with our metric
ansatz~\eqref{eq:g11}, it is useful to introduce new coordinates,
\begin{equation}
\begin{aligned}
   \rho_{a} &=
       -\frac{\sqrt{2}}{e}\mu_{a}X_{a}^{-1/2}\me^{f/2}\me^{(g-f)/n} , \\
   \rho_{i} &=-\frac{\sqrt{2}}{e}\mu_{i}X_{i}^{-1/2}\me^{f/2} ,
\end{aligned}
\label{eq:newcoord}
\end{equation}
where $a=1,\dots,n$ labels the gauged directions and $i=1,..,4-n$ the
ungauged ones. One can then check using the BPS equations that the
frame~\eqref{eq:ortho} can be written as
\begin{equation}
\begin{aligned}
   e^{\rho_{a}} &=\Delta^{-1/6}\me^{-f/2}\me^{(f-g)/n}\dif\rho_{a} , \\
   e^{\phi_{a}} ,
      &= -\Delta^{-1/6}\me^{-f/2}\me^{(f-g)/n}
            \rho_{a}\left(\dif\phi_{a}+2eA_{a}\right) , \\
   e^{\rho_{i}} &= \Delta^{-1/6}\me^{-f/2}\dif\rho_{i} , \\
   e^{\phi_{i}} &= -\Delta^{-1/6}\me^{-f/2}\rho_{i}\dif\phi_{i} .
\end{aligned}
\end{equation}
The $D=11$ metric~\eqref{eq:g11} is then given by 
\begin{equation}
\begin{aligned}
   \dif s_{11}^2 &=  - \Delta^{2/3}\me^{2f}\dif t^2 
         + \Delta^{-1/3}\me^{-f+2(f-g)/n} \sum_{a=1}^{n}
             \left(\dif\rho_a^2+\rho_{a}^2(\dif\phi_a+2e
                 A_{a})^2\right) \\ &\qquad
         + \Delta^{2/3}\me^{2g}\dif s^2(\Sigma) 
         + \Delta^{-1/3}\me^{-f} \sum_{i=1}^{4-n}
             \left(\dif\rho_i^2+\rho_i^2\dif\phi_i^2\right) ,
\end{aligned}
\label{eq:newmetric}
\end{equation}
with the three-form potential
\begin{equation}
   A_{(3)} = - \dif t\wedge \left[
      - \me^{2(f-g)/n}\sum_{a=1}^{n} 
           \rho_{a}\dif\rho_a \wedge (\dif\phi_a+2e A_a)
      + \Delta\me^{f+2g}\Vol(\Sigma) \right] .
\label{eq:A3}
\end{equation}
Comparing with~\eqref{eq:cekt} we can then identify 
\begin{equation}
   H = \Delta^{-1/n}\me^{-3f/n}
\label{eq:Hrel}
\end{equation}
and
\begin{equation}
\begin{aligned}
   g_{AB}\dif y^A \dif y^B &= 
       \Delta^{1/n}\me^{3f/n} \bigg[
          \me^{2(f-g)/n} \sum_{a=1}^{n}\left(
               \dif\rho_a^2+\rho_{a}^2(\dif\phi_a+2e A_{a})^2\right) \\ 
          &\qquad + \Delta\me^{f+2g} \dif s^2(\Sigma) \bigg] .
\end{aligned}
\end{equation}
Given the arguments above that the spatial part of our solutions
describes a complex manifold with a Hermitian metric, we see that
$g_{AB}$ is indeed Hermitian. Since our solution is supersymmetric, preserving
$2^{-n}$ of the supersymmetry for $n=1,2,3,4$ and $2^{-4}$ for $n=5$, we
expect that $H$ and $g_{AB}$ satisfy the conditions given
in~\cite{cekt}. One should note that in general our
ansatz~\eqref{eq:g11} and~\eqref{eq:G4} is not equivalent to the
ansatz~\eqref{eq:cekt}. It was only for the particular BPS solutions
that we were able to rewrite one as the other. 

%%%%%%%%%%%%%%%%%%%%%%%%%%%%%%%%%%%%%%%%

\section{Discussion}

We have presented BPS equations and constructed solutions
of $D=11$ supergravity that are dual to the twisted
theories arising on membranes wrapping holomorphic curves in
Calabi--Yau $n$-folds. For the four-folds and five-folds we found exact 
conformal fixed points when the membrane wraps a Riemann surface
of genus $g>1$. It would be interesting to determine the physical reason 
for such fixed points being present only for these two cases. 
We also analysed the BPS equations numerically and analysed 
the types of singularities encountered in the IR. 
For the five-fold case we managed to find the most general
solutions and it would be nice if the same could be achieved for the
other cases.

In section four we elucidated some of the structure of the solutions by 
presenting the generalised calibrations. In addition a new set of coordinates
was introduced which connects the solutions with other work in the literature.
It is likely that the analysis of this section can be applied to 
other supergravity solutions describing wrapped branes that are obtained
using the technique of \cite{mald}.

For the case of Calabi--Yau five-folds when the scalar fields are vanishing,
we noted that the four-dimensional solution
interpolating from the UV $AdS_4$ region to the
IR $AdS_2\times H_2$ fixed point is in fact the
``topological AdS black hole'' of \cite{caldklemm}. For
the Calabi--Yau 4-folds we numerically demonstrated
a flow from the UV $AdS_4$ region to the
IR $AdS_2\times H_2$ fixed point. This can be similarly considered
to be a ``topological AdS black hole'' with scalar hair. By analogy
with what was found for the five-fold case, it
seems likely that a rotating version also exists.
More generally, it seems likely that the flows to IR AdS fixed points
considered in \cite{malnun,no,agk,gkw,nunpst} will also have rotating
generalisations.

The focus of the paper has been on finding new solutions and exploring
some of their geometry. We now conclude
by briefly discussing the interpretation of the flows from the UV to
the IR from the dual field theory (quantum mechanics) perspective. 
The motion of the
wrapped membranes transverse to the two-cycle and tangent the
Calabi--Yau correspond to possible ``Higgs branches'' while motion
that is also transverse to the Calabi--Yau corresponds to ``Coulomb
branches''. Classically we do not expect Higgs
branches for the case of membranes wrapping the two-sphere as the
corresponding scalar fields of the membrane theory, after twisting,
will not have zero-modes. On the other hand we do expect them for
the case of membranes wrapped on Riemann surfaces with genus $g$
greater than one. Naively then one would expect good singularities in
the IR of the supergravity solutions corresponding to each physical branch. 
For the case of membranes wrapping a two-sphere we thus interpret
the good singularities that arise for the Calabi--Yau two-fold case as
corresponding to the Coulomb branch. For the remaining cases with $l=1$
we only see bad singularities in the IR which suggests that the
Coulomb branches are not accessible in the limits we are considering.
For the $l=-1$ cases we always see a branch of good singularities
which could correspond to either Coulomb or Higgs branches or both. We
expect that any conformal fixed point should appear at the junction
between the two branches. However, for the four-fold case we do have a
fixed point but with good singularities only on one side. This
suggests that in fact again only one branch is accessible in these
solutions.  It would naturally be interesting to investigate the
gravity/field theory correspondence for the flows we have presented
beyond these simple observations. Perhaps the cleanest direction is to
focus on the superconformal quantum mechanics at the IR fixed points
that we found for membranes wrapping Riemann surfaces with $g>1$ in 
Calabi--Yau four-folds and five-folds. 

\subsection*{Acknowledgements}
We thank J. Gutowski for helpful discussions. JPG thanks the EPSRC for
partial support, DJW thanks the Royal Society for support. All authors
are supported in part by PPARC through SPG \#613.

%%%%%%%%%%%%%%%%%%%%

\end{document}